\documentclass[apj,iop]{emulateapj} 
\usepackage{epsfig}
\usepackage{lineno}
\usepackage{epsfig}  
\usepackage{amsmath} 
\usepackage{amssymb}
\usepackage{natbib} 
\usepackage{hyperref}
\usepackage{graphicx} 
\usepackage{endnotes}
\usepackage{color}

\newcommand{\asat}{{\it AstroSat}}

 \shortauthors{Misra et al. }
 \shorttitle{Identification of QPO frequencies of GRS 1915+105}

\begin{document}

\title{Identification of QPO frequency of GRS 1915+105 as the relativistic dynamic frequency of a truncated
accretion disk}

\author{Ranjeev Misra\altaffilmark{1}, Divya Rawat${^*}$\altaffilmark{2}, J S Yadav\altaffilmark{2}, Pankaj Jain\altaffilmark{2} }
\affiliation{Inter-University Center for Astronomy and Astrophysics, Ganeshkhind, Pune 411007, India}
\affiliation{Department of physics, IIT Kanpur, Kanpur, Uttar Pradesh 208016, India}



\email{divyar@iitk.ac.in}

\label{firstpage}

\begin{abstract}
  We have analyzed \asat{} observations of the galactic micro-quasar system GRS 1915+105, when the system exhibited
  C-type Quasi-periodic Oscillations (QPOs) in the frequency range of 3.4-5.4 Hz. The broad band spectra
  (1-50 keV) obtained from simultaneous LAXPC and SXT can be well described by a dominant relativistic
  truncated accretion disk along with thermal Comptonization and reflection. We find that while
  the QPO frequency depends on the inner radii with a large scatter, a much tighter correlation is
  obtained when both the inner radii and accretion  rate of the disk are taken into account. In fact,
  the frequency varies just as the  dynamic frequency (i.e. the inverse of the sound crossing time)
  as predicted decades ago by the relativistic standard accretion disk theory for a black hole with
  spin parameter of $\sim 0.9$. We show that this identification has been possible due to the simultaneous
  broad band spectral coverage with temporal information as obtained from \asat{}.
\end{abstract}

\keywords{accretion, accretion disks --- black hole physics --- X-rays: binaries --- X-rays: individual: GRS 1915+105}

\section{Introduction}\label{intro}
For a test particle orbiting a black hole, there
are three characteristic frequencies depending on the radius \citep{st99a, st99b}. The first is
the Keplerian frequency which is the inverse of the time period of the orbit.
There is the periastron precession frequency which is the Keplerian frequency
minus the Epicyclic one and relates to how an orbit will precess in General Relativity. There is also the
Lense-Thirring precession frequency which is related to the wobbling of the orbit
out of the plane which arises only in General Relativity when the black hole is
spinning.

Apart from these three relativistic test particle frequencies, there are two
other frequencies related to the two characteristic speeds in an accretion disk,
the sound  ($c_s(r)$) and the radial inflow speeds ($v_r(r)$) where $r$ is
 the radial distance.
The dynamical frequency is the inverse of the sound crossing time
i.e. $f_{dyn} \sim c_s(r)/r$. In the standard thin relativistic  accretion disk \citep{sh73,no73} the sound speed
is 
\begin{equation}
c_s(r) = h(r)(GM/r^{3})^{1/2} {\it A^{-1}B^{1}C^{-1/2}D^{1/2}E^{1/2}}
\end{equation} 
where $M$ is the mass of the black hole. The relativistic terms $A$, $B$, $C$, $D$, $E$ are functions of $r$ and the black hole spin patrameter, $a$. They asymptotically tend to unity in the
Newtonian limit i.e. when $r$ tends to infinity.
 The scale height in the inner regions
of the disk is given by
\begin{equation}
  h(r,a) \sim 10^6 {\hbox {cm}}\;\; {\dot M}_{18} {\it A^2B^{-3}C^{1/2}D^{-1}E^{-1}L}
\end{equation}
where ${\dot M_{18}}$ is the accretion rate in units of $10^{18}$ grams/second.
 The relativistic term $L$ is a function of $r$ and $a$ and arises due to the relativistic phenomenon of the existence of
a last stable orbit at which the disk flow no longer depends on viscosity and the height vanishes. Thus
\begin{eqnarray}
  \frac{f_{dyn}}{\dot M_{18}} & =  & N\hspace{0.1cm} 8979 \hspace{0.1cm}{\hbox{Hz}}\; (r/r_g)^{-2.5}\;(M/12.4M_\odot)^{-2} \nonumber \\
  & & \times \;{\it A^1B^{-2}D^{-0.5}E^{-0.5}L}
  \label{fbyM}
\end{eqnarray}
where $r_g = GM/c^2$ is the Gravitational radius and the mass of the black hole,
$M$ has been scaled by $12.4M_\odot$, which is reported black hole mass
for the source GRS 1915+105 \citep{re14}. $N$ is a factor of order unity to incorporate the assumptions made in the standard accretion disk
theory especially in the radiative transfer equation.
 It should be emphasised that A,B,D,E and L
    are functions of radii and are important for small radii, $r < 10 r_g$. Thus
    the functional form of $f_{dyn}$ significantly deviates from its Newtonian
    dependence of $\propto r^{-2.5}$ in this regime.
Note that $f_{dyn}$ does not depend on the unknown
turbulent viscosity parameter $\alpha$ of the standard disk theory in
contrast to the viscous time-scale $\tau_{visc} \sim r/v_r$,
where $v_r$ is the radial inflow velocity of the disk. $\tau_{visc}$ is an order
of magnitude higher than the dynamical time-scale and depends inversely on both
$\alpha$ and accretion rate squared.

X-ray binaries show variability on a wide range of time-scales which include
broad band noise and nearly periodic oscillations termed as Quasi-periodic
Oscillations (QPOs) \citep{va05}. For systems harbouring black holes, the QPO frequency
ranges from milli-Hertz to hundreds of Hertz prompting classification in very low (milli-Hz), low ($~$Hz) and high frequency QPOs ($\sim$100 Hz). Low frequency
QPOs occur at different spectral states and have been further classified
as A, B and C type QPOs \citep{wi99,ho01,re02,ca04}. This multitude of QPOs suggested that they perhaps
correspond to different characteristic timescales of the system described
above. Moreover since the frequency of a particular type of QPO varies,
the radius responsible for the phenomenon should also be varying. An
attractive candidate for this radius is the truncation or inner radius of a
standard disk beyond which there is a hot inner flow \citep{sh76,na08}. Since the
characteristic time-scales depend on General Relativistic corrections,
identification of a QPO frequency with one, opens the exciting possibility
of testing the theory in the strong field regime.

However, as discussed below it has proved to be difficult to make reliable and
independent estimate of the inner disk radius. Indirect schemes have
been employed to identify the QPO frequencies. For example, taking advantage of
the different radial dependencies of the characteristic frequencies, correlation
between frequencies of different QPOs or breaks in the broad band noise have been
used to identify the QPO frequencies \citep{ps99,be02,st99a,st99b}. This method depends on the relatively
rare detection of more than two QPOs at the same time \citep{mo14}. Another method has been
to use the correlation of the QPO frequency with some other features, such as
the high energy spectral index, as proxy for a characteristic radius \citep{ti99}. However,
since the dependence of the high energy spectral index with radius is model dependent and sensitive to assumptions of the unknown viscosity, the best one can
obtain are empirical scaling relations, which have proved useful to compare between different black hole systems \citep{ti04}. 

The inner radius of an accretion disk can be measured by fitting the spectra
of these sources with a truncated accretion disk model \citep{mi99,so00}. Till recently,
the detection of QPOs in black hole systems have been done by the
Proportional Counter Array (PCA) on board the Rossi X-ray Timing Experiment
(RXTE) observatory. Since the radius has to be measured strictly simultaneously
with the QPO, the spectral analysis needed to be restricted to data obtained
from RXTE. However, the PCA had a relatively poor spectral resolution and
its effective energy range was from 3 to 20 keV, while the typical maximum colour
temperature of the disk is around 1 keV. Moreover, since the spectral
data was restricted in the energy range, simple models had to be used
to fit the spectra.  This limited energy range led to severe systematic uncertainties in the
inner disk radius with some values being unphysically small. Moreover, the
results were sometimes contradictory like QPO frequency increasing with
radius for one system while decreasing for another \citep{so00}. Nevertheless, correlations
have been observed between the frequency and the radius which have been used
as evidence for some models, although there were large scatter in the
estimated values \citep{mi09}. A critical limitation of these earlier works was that these analysis were not sensitive enough to test the
  variation of the QPO frequency with accretion rate, since that requires
  broadband data.

The Large Area X-ray Proportional Counter (LAXPC) \citep{ya16a,ag17a} and the Soft X-ray
Telescope (SXT) \citep{si16,si17} on board the Indian Space Observatory AstroSat \citep{ag17b} is ideally
suited to study correlation between the QPO frequency and
the disc inner radius. The high time precision and the large area of LAXPC
provide timing and spectral information in the 4 - 50 keV band, the SXT
provides simultaneous spectral coverage in the low 1.0 - 5.0 keV band.
As reported by \citet{ra19},
AstroSat observed the  black hole system GRS 1915+105 from $28^{th}$ March 2017 18:03:19
till $29^{th}$ March 2017 19:54:07 when the source transited from a relatively steady state called $\chi$ class, through an intermediate state (IMS), to a flaring state (heartbeat state(HS)) where
large amplitude oscillations are seen. All through the observation, the source exhibited C-type QPOs in the frequency range 3.4-5.4 Hz.

In this work, we examine the spectral evolution of the source during this observation and supplement the
results with observations made two days later on $1^{st}$ April 2017, when the source shows both $\chi$ and
'heartbeat state' with QPOs in the same frequency range. It was fortuitous that the source was undergoing
a transition and showed a QPO all through enabling us to study the spectral properties of the source and
correlate them with the varying QPO frequency.

\begin{figure*}
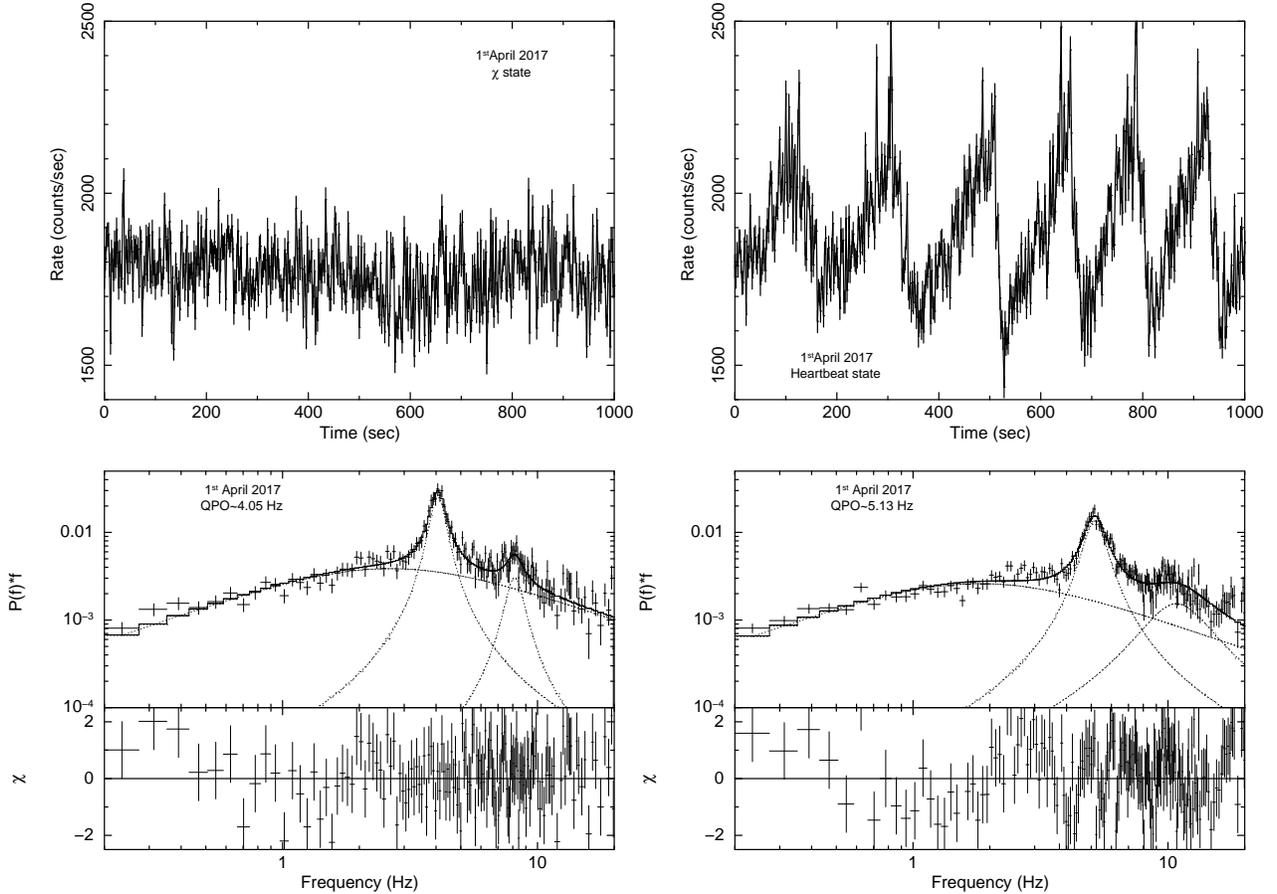

\centering \includegraphics[width=0.33\textwidth,angle=-90]{figures/1st_april_chi.ps}
\centering \includegraphics[width=0.33\textwidth,angle=-90]{figures/heartbeat.ps}
\centering \includegraphics[width=0.33\textwidth,angle=-90]{figures/chi_pds.ps}
\centering \includegraphics[width=0.33\textwidth,angle=-90]{figures/pds_heartbeat.ps}
\caption{Top panel shows the 2.0 sec binned 1000 sec lightcurves of $\chi$ class and heartbeat state. The corresponding PDS in 0.2-20.0 Hz range are shown in bottom panels. LAXPC10 $\&$ LAXPC20 are used for lightcurve and PDS extraction here.}
\label{lightcurve}
\end{figure*}
\begin{figure*}
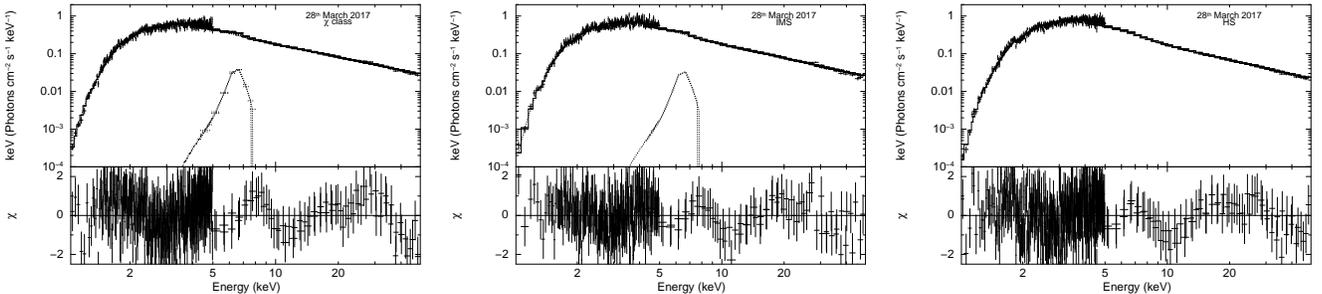

\centering \includegraphics[width=0.22\textwidth,angle=-90]{figures/spectra1.ps}
\centering \includegraphics[width=0.22\textwidth,angle=-90]{figures/spectra2.ps}
\centering \includegraphics[width=0.22\textwidth,angle=-90]{figures/spectra3.ps}
\caption{Spectral fitting including residuals are shown for $\chi$ state, Intermediate state and heartbeat state.}

  \label{spectra}
\end{figure*}


\section{Data Analysis}
\subsection{Timing Analysis}

\cite{ra19} split the $28^{th}$ March 2017 observation into several segments and presented the timing properties for each
of them. From the Power Density Spectra (PDS) they obtained the QPO frequencies for segments corresponding to
the $\chi$, intermediate and heartbeat states. Following \cite{ra19} we do the same analysis for the
\asat{} LAXPC observation of GRS 1915+105 during $1^{st}$ April 2017 00:03:21 till $1^{st}$
April 14:38:01. The data was analyzed for the two  units LAXPC 10 and LAXPC 20, using the LaxpcSoft\footnote{\url {http://astrosat-ssc.iucaa.in/?q=laxpcData}\label{note1}}.  Like the earlier observation, the source exhibited $\chi$ and heartbeat states which were divided
into 6 segments (2 for $\chi$ and 4 for the heartbeat state). Lightcurves for representative segments for
a $\chi$ and heartbeat state are shown in top panel of Figure \ref{lightcurve} and the corresponding PDS
are shown in the bottom panel. The PDS were fitted using lorentzian functions and the QPO frequency with
error was estimated using the same technique given in \cite{ra19}. Thus combining the two observations
we have a total of 16 segments ( 5 for $\chi$, 3 for intermediate and 8 for the heartbeat states) for
which the QPO frequency has been estimated and tabulated in the first column of Table \ref{specpar}.

\subsection{Spectral Analysis}

For each of the 16 segments, simultaneous spectral data was obtained from LAXPC 10, 20 and SXT.
The LAXPC spectra, background and response files were generated using the LaxpcSoft\footnotemark[\value{footnote}]. For SXT data reduction recent arf and rmf files are used, details of which are given at \asat{} website \footnote{\label{note2}\url {http://astrosat-ssc.iucaa.in/?q=sxtData}}.
The SXT spectra were extracted from a source region of 12 
arcmins and the standard background spectra were
used for all spectra.

The SXT (energy range 1.0 - 5.0 keV) and LAXPC 10 and 20 (
energy range 4 - 50 keV) spectra of each data
set was analyzed together 
using the X-ray spectra fitting software XSPEC \citep{ar96} using its inbuilt models.
During the spectral fitting gain variation for SXT was taken into account by using the
gain fit command in XSPEC. The offset value obtained ranged from 1.4 to 2.4 eV. Additional systematic error of 
3\% was included. To take into account possible uncertainties in the effective area of the
instruments a variable constant was included to the LAXPC 10 and 20 spectra relative to SXT, whose values
ranged from 0.81 to 0.92.

The spectra were fitted using the relativistic disk model,
``kerrd'' \citep{eb03}, and the convolution model ``simpl'' \citep{st09} to take into account the
Comptonization of the disk photons in the inner flow. The accretion rate
and the inner radius of the disk were estimated from the best fit values
obtained from the ``kerrd'' model. The mass of the black hole, distance to
the source and inclination angle of the disk were taken to be $12.4 M_\odot$ ,
$8.6$ kpc and $60^o$ \citep{re14} respectively. The colour factor was fixed to $1.7$
\citep{sh95}.

To take into account a relativistically
smeared Iron fluorescence line, the model ``kerrdisk'' \citep{br06} was used. While
the ``kerrd'' model implicitly assumes a fast spinning black hole,
the spin is a parameter for ``kerrdisk'' which was fixed at $0.98$ \citep{bl09}, as the
spectral fitting was found to be insensitive to its value. For the kerrdisk the  emissivity index for both the inner and
outer parts of the disk was fixed at 1.8 \citep{bl09}. The rest frame
energy  of the iron line was fixed at $6.4$ keV \citep{bl09}. The inner radius  was tied to that used
for the ``kerrd'' after dividing by an appropriate factor of 1.235,
since for ``kerrd'', the radius is measured in $r_g$, while for the kerrdisk
it is in units of the radius of marginal stability.
Absorption by
intervening matter was modelled using ``tbabs'' \citep{wi00} with a column density fixed at $4\times 10^{22} cm^{-2}$\citep{bl09}.
Representative spectra for a  $\chi$, intermediate  and heartbeat state are shown in figure \ref{spectra}. Note that
for the spectra of heartbeat state the iron line component is insignificant. 
Table \ref{specpar} lists the best fit values of the parameters which are the accretion rate,
inner disk radius, the fraction scattered into the Comptonizing medium,
the index of the Comptonized spectrum and flux in the line emission.

\begin{table*}
 \centering
 \caption{Spectral Parameters for GRS 1915+105 in 1.0-50.0 keV energy range}
\begin{center}
\scalebox{0.9}{%
\begin{tabular}{ccccccccc}
\hline  
Exposure time & State &QPO frequency& Accretion rate & Inner radius & Fraction scatter&  Gamma & Flux in line emission &  $\chi^{2}/$Dof\\  
(sec)  & & (Hz) & $10^{18}$gm s$^{-1}$  & (R$_{g}$) &  &  &  $10^{-2}$photons cm$^{-2}$s$^{-1}$&  \\ 
\hline
 & & & &28$^{th} March$ 2017 & & & &\\
1199 & $\chi$ class & $3.59_{-0.01}^{+0.01}$ & $0.67_{-0.02}^{+0.01}$ & $4.62_{-0.06}^{+0.15}$ & $0.42_{-0.01}^{+0.01}$ & $2.169_{-0.005}^{+0.005}$ & $1.1_{-0.1}^{+0.2}$ &  488.7/426\\
1199 & $\chi$ class & $3.46_{-0.02}^{+0.02}$ & $0.74_{-0.03}^{+0.04}$ & $5.25_{-0.24}^{+0.27}$ & $0.43_{-0.02}^{+0.02}$ & $2.169_{-0.005}^{+0.011}$ & $1.1_{-0.2}^{+0.2}$ &  492.1/424\\
1203 & $\chi$ class & $3.65_{-0.02}^{+0.01}$ & $0.77_{-0.03}^{+0.03}$ & $5.31_{-0.15}^{+0.19}$ & $0.45_{-0.01}^{+0.02}$ & $2.221_{-0.007}^{+0.006}$& $1.0_{-0.2}^{+0.2}$ &  509.4/431\\
903 & $IMS$ & $4.08_{-0.02}^{+0.03}$ & $0.71_{-0.04}^{+0.03}$ & $4.20_{-0.29}^{+0.20}$ & $0.39_{-0.01}^{+0.01}$ & $2.223_{-0.011}^{+0.009}$ & $0.9_{-0.2}^{+0.3}$ &  489.4/411\\
477 & $IMS$ & $4.17_{-0.03}^{+0.03}$ &$0.74_{-0.04}^{+0.04}$ & $4.53_{-0.30}^{+0.25}$ & $0.40_{-0.01}^{+0.02}$ & $2.247_{-0.013}^{+0.012}$ & $0.9_{-0.3}^{+0.3}$ & 298.8/322\\
840 & $IMS$ & $4.38_{-0.06}^{+0.07}$ & $0.68_{-0.04}^{+0.02}$ & $3.58_{-0.27}^{+0.17}$ & $0.35_{-0.01}^{+0.01}$ & $2.260_{-0.014}^{+0.013}$ & -- &  419.1/410\\
1209 & $ HS$ & $5.08_{-0.03}^{+0.04}$ & $0.65_{-0.03}^{+0.01}$ & $3.02_{-0.24}^{+0.09}$ & $0.29_{-0.01}^{+0.01}$ & $2.255_{-0.017}^{+0.014}$ & -- &  519.6/437\\
1211 & $ HS$ & $5.15_{-0.03}^{+0.03}$ & $0.64_{-0.00}^{+0.03}$ & $2.90_{-0.03}^{+0.18}$ & $0.29_{-0.01}^{+0.01}$ & $2.241_{-0.009}^{+0.014}$ & -- &  492.7/436\\
1213 & $ HS$ & $5.33_{-0.03}^{+0.03}$ &$0.66_{-0.02}^{+0.02}$ & $2.80_{-0.09}^{+0.12}$ & $0.27_{-0.01}^{+0.01}$ & $2.239_{-0.006}^{+0.012}$ & -- &  539.4/439\\
1216 & $HS$ & $5.42_{-0.03}^{+0.03}$ & $0.69_{-0.01}^{+0.01}$ & $3.03_{-0.14}^{+0.09}$ & $0.26_{-0.01}^{+0.01}$ & $2.252_{-0.007}^{+0.012}$ & -- &  490.5/437\\
& & & &$1^{st} April$ 2017 & & & & \\
634 & $\chi$ class & $4.05_{-0.02}^{+0.02}$ & $0.78_{-0.04}^{+0.04}$ & $4.84_{-0.31}^{+0.32}$ & $0.38_{-0.01}^{+0.01}$ & $2.219_{-0.012}^{+0.012}$ & $1.0_{-0.2}^{+0.2}$ &  347.2/374\\
204 & $\chi$ class & $4.67_{-0.04}^{+0.04}$ & $0.79_{-0.07}^{+0.06}$ & $4.42_{-0.47}^{+0.35}$ & $0.36_{-0.01}^{+0.01}$ & $2.274_{-0.020}^{+0.017}$ & $0.7_{-0.2}^{+0.4}$ &  222.3/199\\
748 & $HS$ & $5.13_{-0.03}^{+0.03}$ & $0.72_{-0.02}^{+0.02}$ & $3.44_{-0.20}^{+0.12}$ & $0.31_{-0.00}^{+0.00}$ & $2.278_{-0.003}^{+0.003}$ & -- &  438.1/407\\
1175 & $HS$ & $5.13_{-0.02}^{+0.02}$ & $0.71_{-0.03}^{+0.02}$ & $3.27_{-0.17}^{+0.13}$ & $0.31_{-0.01}^{+0.01}$ & $2.287_{-0.014}^{+0.015}$ & -- &  519.5/439\\
1260 & $HS$ & $5.12_{-0.01}^{+0.01}$ & $0.72_{-0.02}^{+0.02}$ & $3.44_{-0.13}^{+0.14}$ & $0.30_{-0.01}^{+0.01}$ & $2.262_{-0.013}^{+0.013}$ & -- &  458.9/439\\
762 & $HS$ & $5.27_{-0.04}^{+0.04}$ & $0.69_{-0.01}^{+0.04}$ &$3.09_{-0.10}^{+0.22}$ & $0.28_{-0.01}^{+0.01}$ & $2.245_{-0.014}^{+0.014}$ & -- &  459.4/411\\
\hline
\end{tabular}}
\tablecomments{Here, IMS and HS stand for intermediate and Heartbeat states respectively. }
\end{center}
\label{specpar}
\end{table*}

\begin{figure*}
\centering \includegraphics[width=0.33\textwidth,angle=-90]{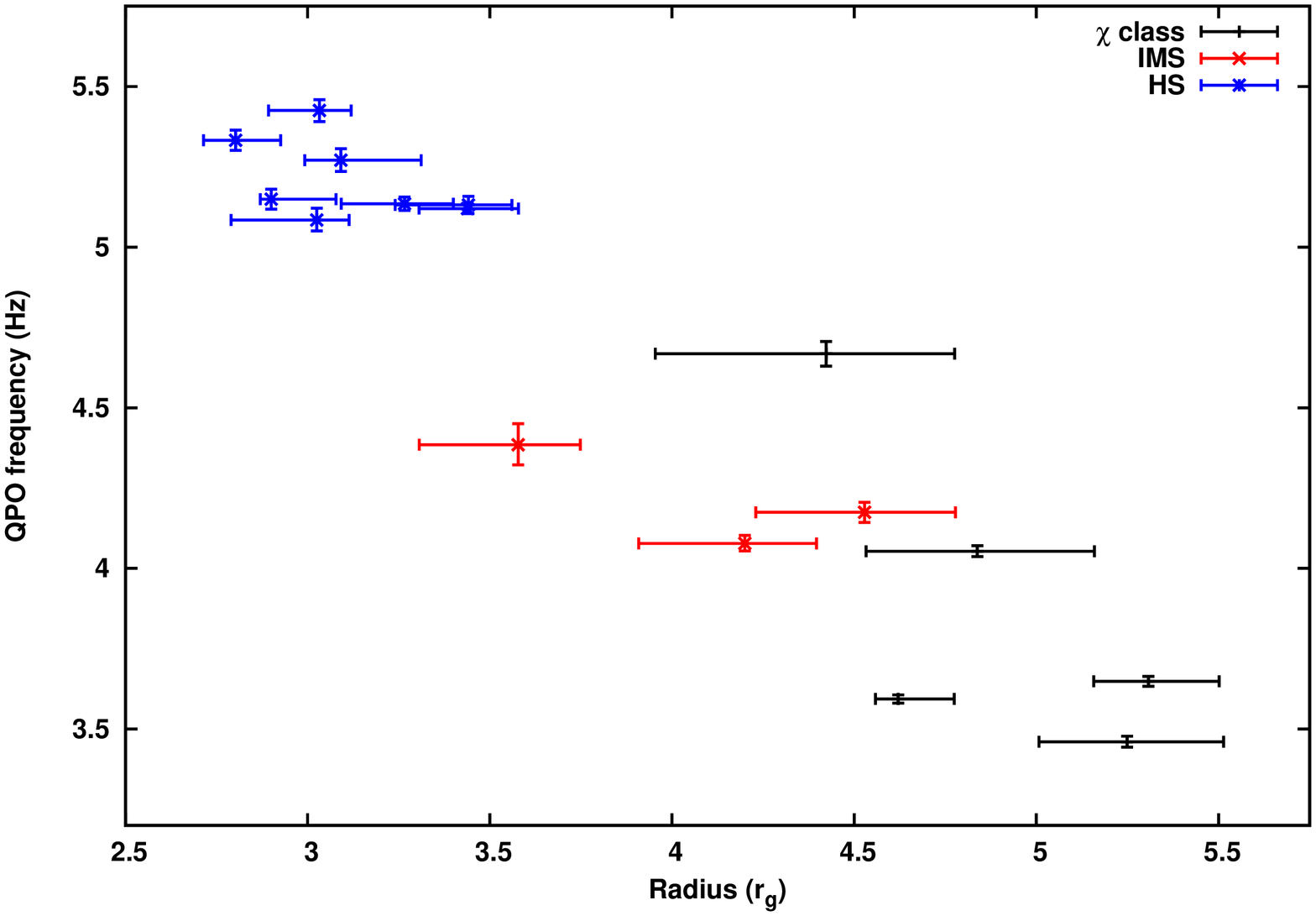}
\centering \includegraphics[width=0.33\textwidth,angle=-90]{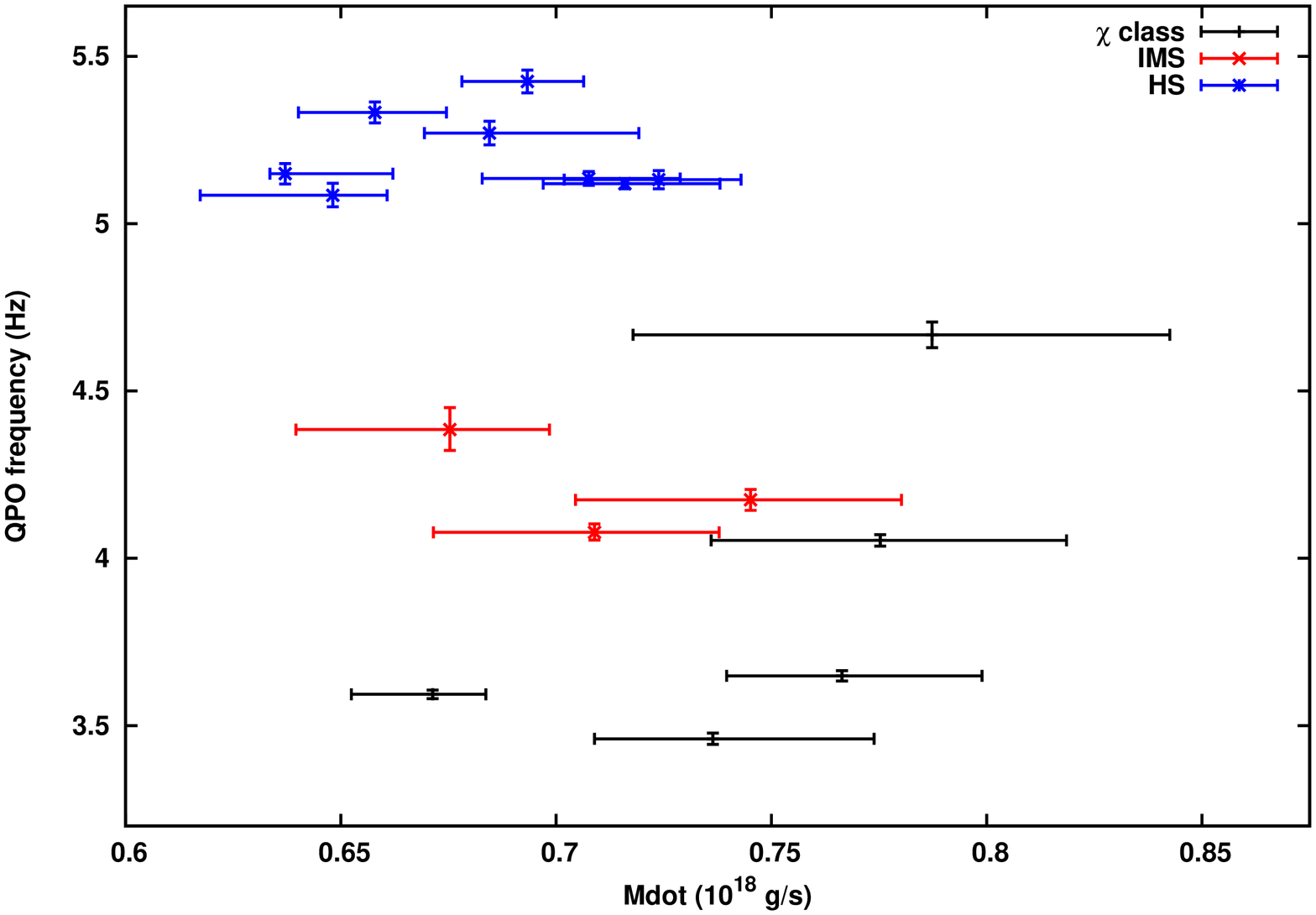}
\centering \includegraphics[width=0.33\textwidth,angle=-90]{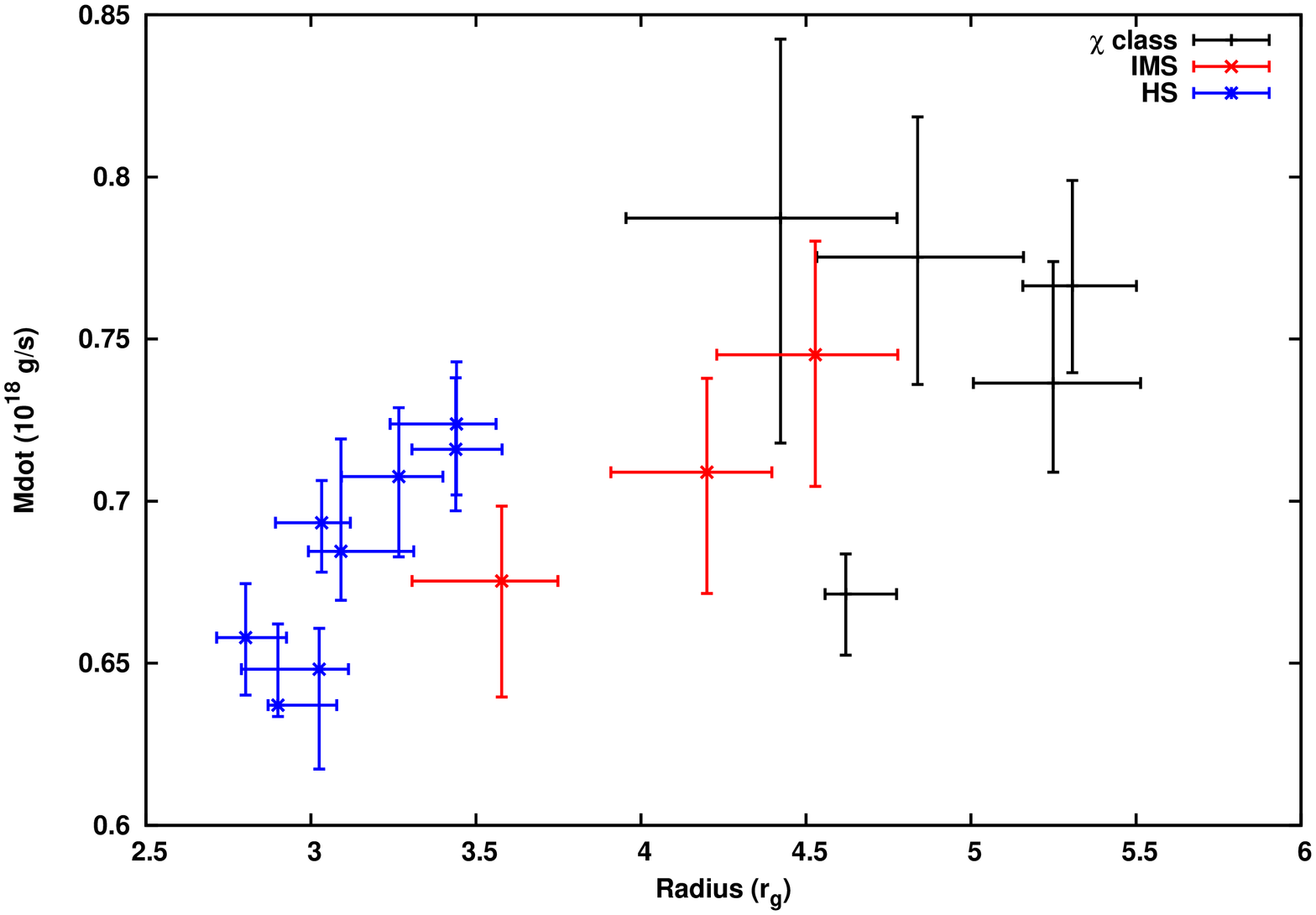}
\caption{The variation of QPO frequencies with inner disk radii and Accretion rate are shown in upper left and upper right panels respectively.  The bottom panel shows the variation of the accretion rate with inner disk radii. }
\label{Figure1}
\end{figure*}

\begin{figure*}
\centering \includegraphics[width=0.33\textwidth,angle=-90]{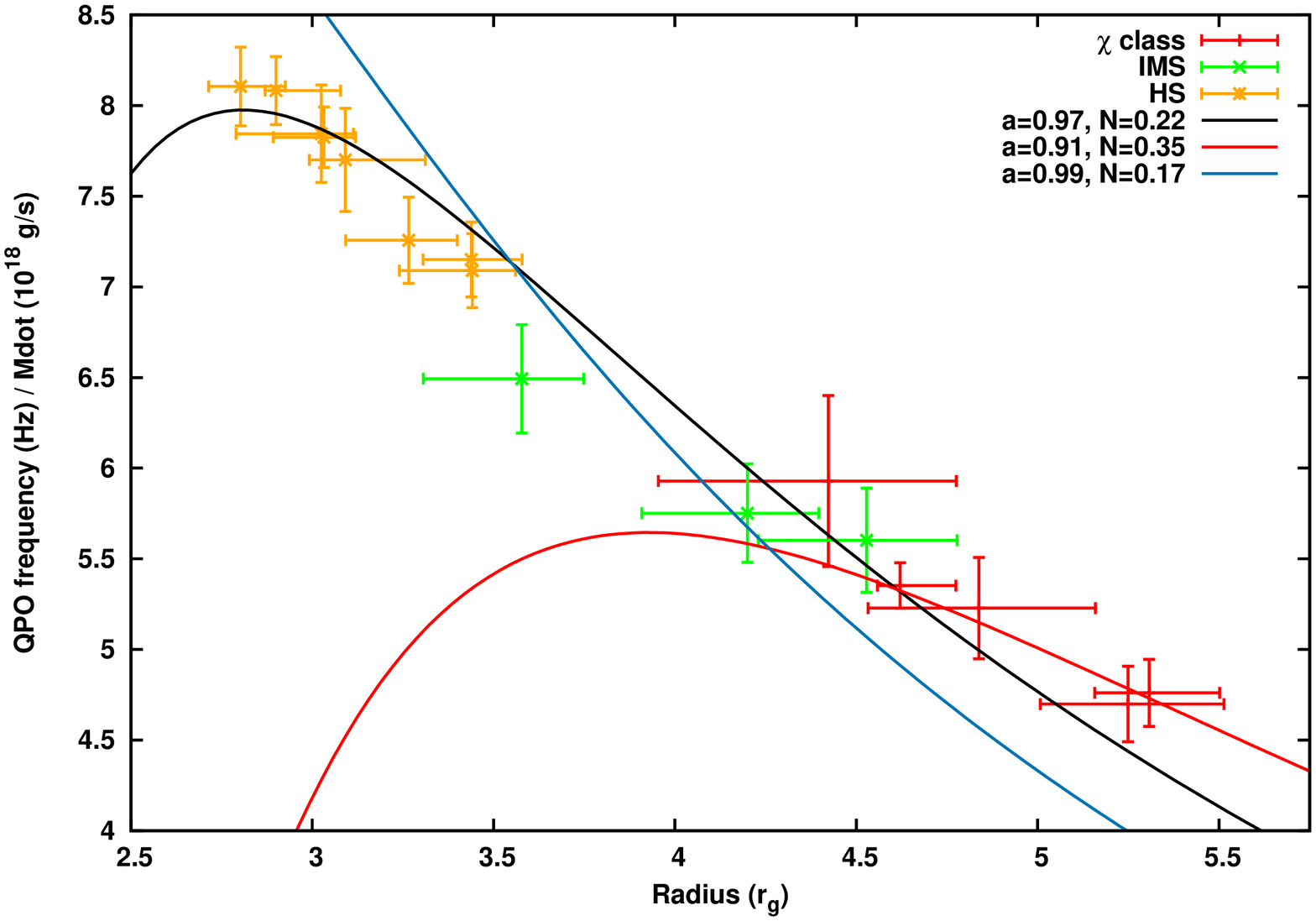}
\centering \includegraphics[width=0.33\textwidth,angle=-90]{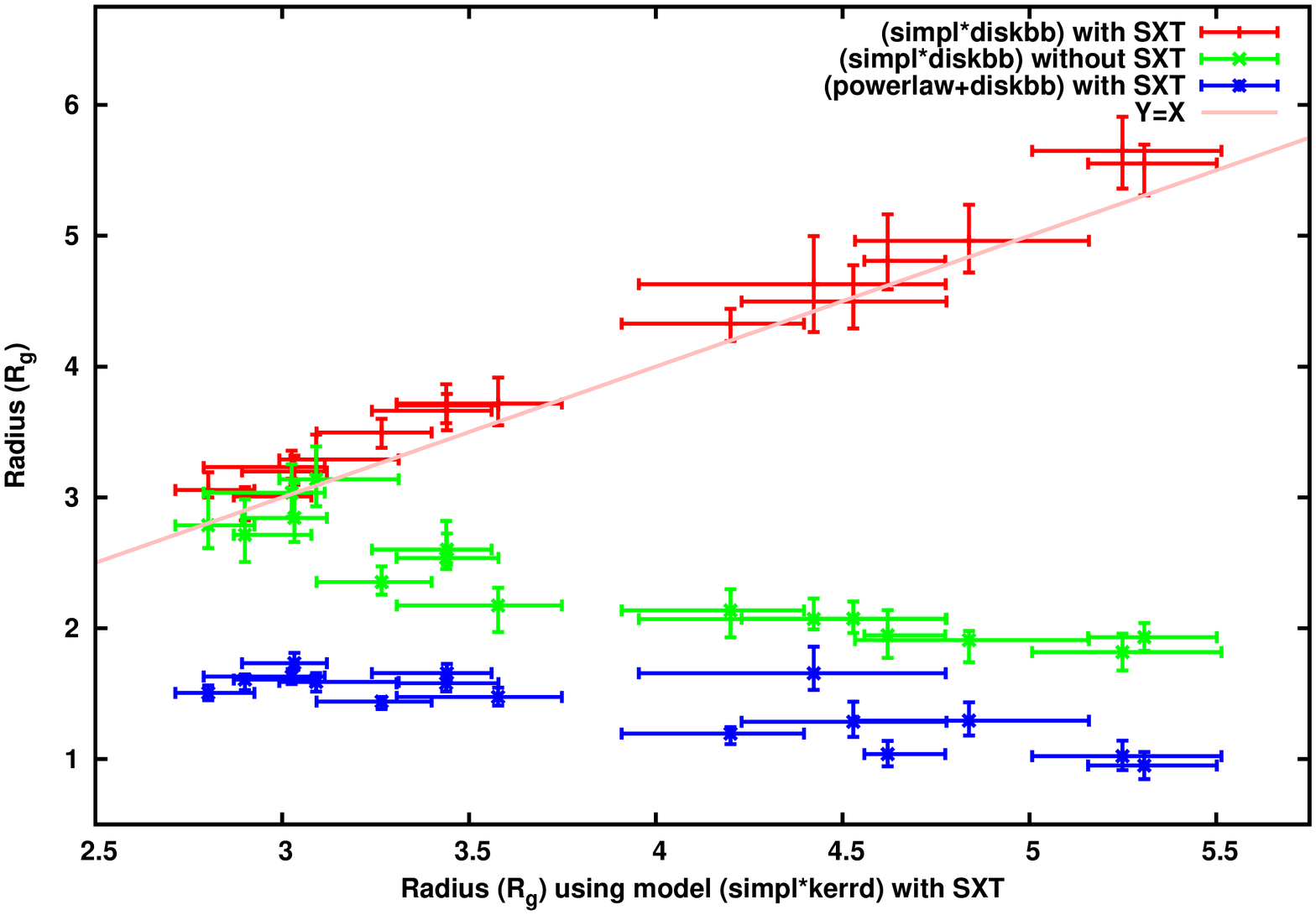}
  \caption{Left panel shows variation of QPO frequency divided by the accretion rate with inner disk radii. The lines represent the
    $\frac{f_{dyn}}{\dot M_{18}}$ as predicted by the relativistic standard accretion disk model (Equation \ref{fbyM}) for
    dimensionless spin parameter $a = 0.973 \pm 0.002$ (best fit with $N = 0.22 \pm 0.01$ and reduced $\chi^{2}$ $\sim$0.5), $a = 0.91$ ($N = 0.35$) and $a = 0.99$ ($N = 0.17$). In the right panel we show a
comparison of the estimated values of the inner disk radii.}
  \label{radcomp}
\end{figure*}

\section{Results}

The upper left panel of Figure \ref{Figure1} shows the variation of the QPO frequency with radius
where a broad anti-correlation is visible, however, it is difficult to quantify the
dependence because of the  significant scatter. Indeed, the scatter suggests that the QPO frequency depends
not only on the inner radius but also on some other parameter. The upper right and bottom panels of Figure \ref{Figure1} show the
  variation of the frequency with the accretion rate and the accretion rate
  with inner radii , where again there seems to be a correlation but with a large scatter.
However, if one considers the frequency to depend both on the radius and the accretion rate and in particular
if it is of the form $\propto {\dot M} F(R_{in})$ then the correlation is significantly better. This
is illustrated in left panel of Figure \ref{radcomp} where the QPO frequency divided by the accretion rate is plotted
against the inner radius.  More pertinently the variation is the same as predicted by the
standard accretion model for the dynamic frequency (Equation \ref{fbyM})
represented by lines for different values of the spin parameter $a$. Note that the predicted functional
form depends only on $a$ and the normalization factor $N$ which should be of order unity.
While a formal fit gives $a = 0.973 \pm 0.002$, we also show the variation
for two different values of $a=.91$ and $a=.99$ to illustrate the
constraints the data imposes on a.

It is interesting to note that the best fit value of the black hole spin parameter obtained here is $a \sim 0.973 \pm 0.002$ which is consistent with $a = 0.98 \pm 0.01$ obtained independently by fitting the relativistically blurred reflection model  to the broad band spectrum from Suzaku  \citep{bl09}.  While earlier results using RXTE and ASCA spectra gave contradictory results \citep{mc06,mi06}, the better spectral resultion of Suzaku and broad band analysis makes the results obtained by \citet{bl09} more reliable. We stress that the consistent determination of the spin parameter using two completely independent different methods,  significantly strengthens the interpretation present in this work.

We emphasis that primary result used in this work, i.e. the
estimate of the inner disk radii is not sensitive to the relativistic
``kerrd'' and ``kerrdisk'' models and they have been invoked for
consistency. An alternate empirical model ``tbabs*(simpl*diskbb+Gaussian)''
provides nearly the same estimates of the inner radii as shown
in right panel of Figure \ref{radcomp} where the two radii estimates are compared. There are two primary reasons for obtaining
a reliable value of the radii which are (a) the presence of low energy
data from SXT and (b) the use of ``simpl'' to model the Comptonization
instead of a power-law. This is illustrated in right panel of Figure \ref{radcomp},
where the radii estimates without the SXT data (using the same empirical
model  (``tbabs*(simpl*diskbb+Gaussian)'') is plotted against that obtained
from the relativistic models with SXT. Without SXT data the radii estimated
are systematically lower and not correlated with the ones obtained when
SXT data is considered.  The right panel of Figure \ref{radcomp} also shows the case, when
a power-law model is used instead of ``simpl'' (with SXT data). In this case also
the radii obtained is systematically under estimated and not well correlated with
the values estimated when ``simpl'' is used.

During the heartbeat state, the overall flux and spectra evolves
    \citep{ra19}, while here we have considered a time-averaged spectrum. 
    To verify the impact of this on the results presented, we performed flux
    resolved spectroscopy for all heartbeat observations by dividing the data
    into three flux levels and obtaining the corresponding spectra. We find that
    the qualitative nature as shown by the left panel of Figure 4 does not change with best fit values, a = $0.968\pm 0.002$ and N = $0.24\pm 0.01$ with reduced $\chi^{2}$ $\sim$1.1, close to
    the ones obtained using time-averaged spectra.


\section{Discussion and Summary}

The result obtained in this paper relies on the accuracy of some measured  and theoretically
estimated quantities. Future improvement on the estimate of these would
refine the fitting presented here and can provide a robust value of the
spin parameter.
 These include the uncertainties in (a)  the estimated distance to the
source, mass of the black hole and the inclination angle used; (b) the
effective area and response of the LAXPC and SXT detectors,  and (c)
the theoretically estimated
colour factor, especially since this was done for a non-spinning black hole
\citep{sh95}.
Note that most of these uncertainties are independent of each other and
will give rise to a secular shift in the radii and accretion rate.

 The analysis has been done by fixing the neutral column density value,
    nH at $4\times 10^{22} cm^{-2}$ as obtained by Blum et al 2009 using Suzaku data.
    If we instead allow it to vary its value ranges from  $3.5 \times 10^{22} cm^{-2}$ to $4 \times 10^{22} cm^{-2}$ for different orbits with a typical error of $0.1 \times 10^{22} cm^{-2}$. Since we expect the column density not to vary during the course of the observation we have used the value obtained
    by \citet{bl09}. If instead we use the average value
    obtained from the present observation, i.e. we fix it
    to  $3.75 \times 10^{22} cm^{-2}$, we get qualitatively similar results,
    with  the best fit values for the
    spin parameter and normalization to be  a = $0.986\pm 0.002$ and N = $0.18\pm 0.01$.

It is interesting to note that for small values of radius (i.e. $r \sim 4 r_g$) the  radial functional
form of Equation \ref{fbyM}, is approximately $1/r$ instead of 1/$r^{2.5}$ due to its dependence
on the relativistic terms, A,B,D,E and L. This means that the QPO frequency is roughly
proportional to ${\dot M}/R_{in}$, which in turn is proportional to the  disk flux. Thus, an approximate
dependence of the QPO frequency on total flux (if the disk component dominates) is expected in this scenario.
Moreover, the spectral index of the Comptonization component may also depends on the disk flux, leading to a QPO
frequency dependence on the index. For the spectral fitting results presented here, there is indeed a dependence
of the spectral index on the disk flux, which will be  physically interpreted in a later work 
where the evolution of the spectral parameters will be described in more detail. Here, we note, that in this
interpretation dependence of QPO frequency on spectral index \citep{bh19} is an indirect consequence of it being the dynamical frequency of a truncated disk as explictly mentioned in \citet{ti99}.

Since the QPO frequency depends both on the radius and accretion rate, this
 favours
models based on hydrodynamics, especially for example the coronal oscillatory
one \citep{ti99,ti04,sh07} where the frequency is indeed identified with the dynamic one. However,
other models such as the Accretion-Ejection Instability \citep{va02,ta99} are also promising,
since evidence is provided for a driving instability. Although some of these
theoretical models have different identification of the QPO frequency, the
result presented here provides now a strong foothold, on which sophisticated
models can be developed.

In summary, we exploit the broad band capability of \asat{} to study the spectral
properties of GRS 1915+105 with the QPO frequency. We find that the frequency
depends on the accretion rate and inner radius of the disk, just as it was predicted for the dynamical frequency of a relativistic accretion disk. Thus, we identify the QPO frequency as the inverse of the sound cross time from the inner
disk radius where strong General Relativistic effects dominate.

\section{Acknowledgment}
We thank the referee for constructive comments. This research has used the data of AstroSat mission of the Indian Space Research Organisation (ISRO), archived at the Indian Space Science Data Centre. The authors would like to acknowledge the support from the LAXPC Payload Operation Center (POC) and SXT POC at the TIFR, Mumbai for providing support in data reduction.
\bibliography{ms}
\end{document}